\documentclass[twocolumn]{aastex631}

\usepackage{amsmath}
\usepackage{xspace}
\usepackage{hyperref}

\newcommand{\nustar}{\textit{NuSTAR}\xspace}
\newcommand{\swift}{\textit{Swift}\xspace}
\newcommand{\nicer}{\textit{NICER}\xspace}
\newcommand{\fermi}{\textit{Fermi}\xspace}
\newcommand{\atca}{ATCA\xspace}
\newcommand{\neowise}{\textit{NEOWISE}\xspace}
\newcommand{\igr}{IGR~J17091--3624\xspace}
\newcommand{\flux}{erg\,cm$^{-2}$\,s$^{-1}$\xspace}

\begin{document}

\title{Unraveling the hybrid origins of the X-ray non-thermal emission from IGR~J17091--3624}

\correspondingauthor{Yanan Wang}
\email{wangyn@bao.ac.cn}

\author[0000-0001-9576-1870]{Zikun Lin}
\affiliation{Key Laboratory of Optical Astronomy, National Astronomical Observatories, Chinese Academy of Sciences, Beijing 100101, People’s Republic of China}
\affiliation{School of Astronomy and Space Sciences, University of Chinese Academy of Sciences, Beijing 100049, People’s Republic of China}

\author[0000-0003-3207-5237]{Yanan Wang}
\affiliation{Key Laboratory of Optical Astronomy, National Astronomical Observatories, Chinese Academy of Sciences, Beijing 100101, People’s Republic of China}

\author[0000-0002-5761-2417]{Santiago del Palacio}
\affiliation{Department of Space, Earth and Environment, Chalmers University of Technology, SE-412 96 Gothenburg, Sweden.}

\author[0000-0003-2187-2708]{Mariano M\'endez}
\affiliation{Kapteyn Astronomical Institute, University of Groningen, PO Box 800, CH-9700 AV Groningen, the Netherlands}

\author[0000-0001-5586-1017]{Shuang-Nan Zhang}
\affiliation{Key Laboratory of Particle Astrophysics, Institute of High Energy Physics, Chinese Academy of Sciences, 19B Yuquan Road, Beijing 100049, People’s Republic of China}
\affiliation{School of Astronomy and Space Sciences, University of Chinese Academy of Sciences, Beijing 100049, People’s Republic of China}

\author[0000-0002-7930-2276]{Thomas D. Russell}
\affiliation{INAF, Istituto di Astrofisica Spaziale e Fisica Cosmica, via Ugo la Malfa 153, 90146 Palermo, Italy}

\author[0000-0001-9599-7285]{Long Ji}
\affiliation{School of Physics and Astronomy, Sun Yat-sen University, Zhuhai 519082, People’s Republic of China}

\author[0000-0003-3554-2996]{Jin Zhang}
\affiliation{School of Physics, Beijing Institute of Technology, Beijing 100081, People’s Republic of China}

\author[0000-0003-4498-9925]{Liang Zhang}
\affiliation{Key Laboratory of Particle Astrophysics, Institute of High Energy Physics, Chinese Academy of Sciences, 19B Yuquan Road, Beijing 100049, People’s Republic of China}

\author[0000-0002-3422-0074]{Diego Altamirano}
\affiliation{School of Physics and Astronomy, University of Southampton, Southampton, Hampshire SO17 1BJ, UK}

\author{Jifeng Liu}
\affiliation{Key Laboratory of Optical Astronomy, National Astronomical Observatories, Chinese Academy of Sciences, Beijing 100101, People’s Republic of China}
\affiliation{School of Astronomy and Space Sciences, University of Chinese Academy of Sciences, Beijing 100049, People’s Republic of China}
\affiliation{Institute for Frontiers in Astronomy and Astrophysics, Beijing Normal University, Beijing 102206, People’s Republic of China}
\affiliation{New Cornerstone Science Laboratory, National Astronomical Observatories, Chinese Academy of Sciences, Beijing 100012, People’s Republic of China}

\begin{abstract}

We present a comprehensive study based on multi-wavelength observations from the \nustar, \nicer, \swift, \fermi, \neowise, and \atca telescopes during the 2022 outburst of the black hole X-ray binary \igr. Our investigation concentrates on the heartbeat-like variability in the X-ray emission, with the aim of using it as a tool to unravel the origin of the non-thermal emission during the heartbeat state. Through X-ray timing and spectral analysis, we observe that the heartbeat-like variability correlates with changes in the disk temperature, supporting the disk radiation pressure instability scenario. Moreover, in addition to a Comptonization component, our time-averaged and phase-resolved spectroscopy reveal the presence of a power-law component that varies independently from the disk component. Combined with the radio to X-ray spectral energy distribution fitting, our results suggest that the power-law component could originate from synchrotron self-Compton radiation in the jet, which requires a strong magnetic field of about $B = (0.3$--$3.5)\times10^6$\,G. 
Additionally, assuming that \igr and GRS~1915+105 share the same radio--X-ray correlation coefficient during both the hard and the heartbeat states, we obtain a distance of $13.7\pm2.3$\,kpc for \igr.

\end{abstract}


\section{Introduction} \label{sec:intro}

Black hole X-ray binaries (BHXRBs) are a type of celestial systems that consist of a black hole in a close orbit, accreting material from its companion star, and show repeated outbursts with a cadence of years. During outbursts, the radiation of BHXRBs consists of thermal and/or non-thermal components. The thermal component is typically represented by a multi-color blackbody emission, emanating from the accretion disk in the soft X-ray band \citep[see e.g.,][]{Shakura1973}. The non-thermal emissions in the X-ray band are particularly complex, primarily attributable to processes such as Comptonization and synchrotron radiation occurring in different regions, which make them as enduring enigmas in high-energy astrophysics \citep[see e.g.,][]{Illarionov1975,Rybicki1979}. 

The non-thermal X-ray emission from BHXRBs has been generally believed to be dominated by a hot, optically thin electron gas cloud -- commonly termed as the `corona' -- which undergoes inverse Compton scattering with seed photons from the accretion disk, resulting in a (cutoff) power-law spectrum \citep[see e.g.,][]{Sunyaev1980,Haardt1993,Gilfanov2010,GarciaFederico2021,GarciaFederico2022}.
Photons reflecting back to the disk from the corona produce an iron ${\rm K\alpha}$ emission line, peaking at 6.4--6.9\,keV, which is broadened by Doppler and relativistic effects along with a Compton hump peaking around 10--30\,keV \citep[see e.g.,][]{Lightman1980ApJ...236..928L,Lightman1981ApJ...248..738L,Fabian1989,Dauser2010,Garcia2014,Dauser2022}.
On the other hand, jets have been indicated to emit a broad spectrum ranging from radio to gamma rays through synchrotron and synchrotron self-Compton (SSC) radiation. The emission at high energies displays a power-law spectrum in the optically thin regime, represented as \( F_\nu \propto \nu^{\alpha} \) \citep[see e.g.,][]{Marscher1983,Sari1998,Sari2001,Finke2008,Nakar2011,Hoshino2013,Ball2018}. Despite numerous efforts, differentiating these emission regions between corona and jets using only X-ray data has been unsuccessful \citep[e.g.,][]{Heinz2004,Markoff2005}. However, more recent approaches using varied techniques, i.e. well defined optical/near-infrared jet emission and polarization measurement, have successfully demonstrated that non-thermal X-ray emission may sometimes originate from jets \citep[e.g.,][]{RussellDM2010,RussellDM2014,Chattopadhyay2024}. Furthermore, broadband spectral energy distribution (SED) fitting has emerged as an effective method for distinguishing non-thermal emissions. \citep[e.g.,][]{RussellDM2014,Punsly2011,Kantzas2021,Rodi2021}.

Timing analysis, which involves studying the variability in the lightcurve across different energy bands, also aids in our understanding of the correlation between the disk, corona, and jet \citep[e.g.,][]{Wang2021,Mariano2022}. As the jet moves perpendicularly away from the accretion disk with relativistic speed \citep[e.g.,][]{Pushkarev2009,Pushkarev2017,Tetarenko2017,Miller2019,Zdziarski2022}, the scattering efficiency between disk photons and jet electrons significantly decreases compared with the disk and corona \citep[e.g.,][]{Wilkins2012}. Long-term variability (lasting hours to days) between the disk and jet could remain correlated \citep[e.g.,][]{Fender2004,Vincentelli2023,You2023}, but short-term oscillations (lasting several seconds) originating from the disk are expected to dilute during the propagation \citep[e.g.,][]{Eikenberry1998}. Thus, integrating timing analysis of short-term variability from the disk with broadband SED spectroscopy could elucidate the radiation mechanisms in BHXRBs.

Heartbeat variability in BHXRBs, characterized by quasi-periodic flares in X-ray emission, is suggested to stem from the radiation pressure instability of accretion disk \citep[e.g.,][]{Janiuk2000,Nayakshin2000,Neilsen2011,Neilsen2012,Yan2017}.
Such a phenomenon is associated with a limit-cycle behavior in the accretion rate and disk surface density, resulting in a continuous evacuation and refilling of the inner accretion disk  \citep[e.g.,][]{Abramowicz1988,Belloni1997,Merloni2006,Pan2022}. 
Currently, it has only been observed in two BHXRBs, GRS~1915+105 (hereafter, GRS~1915, \citealt{Belloni1997,Neilsen2011}) and \igr (hereafter, IGR~J17091, \citealt{Altamirano2011,Court2017MNRAS.468.4748C}). Other types of accreting systems have also been observed to exhibit similar quasi-periodic flares, which are related to disk behaviors, such as the ultraluminous X-ray source, NGC~3621 \citep{Motta2020}, and one neutron star X-ray binary, Swift~J1858.6--0814 \citep{Vincentelli2023}.

IGR~J17091 is a galactic low-mass BHXRB discovered with {\it INTEGRAL/IBIS} in April 2003 \citep{Kuulkers2003}. Later on, IGR~J17091 had experienced another four typical outbursts in 2007, 2011, 2016 and 2022 \citep{Capitanio2009,Krimm2011,Miller2016,Miller2022}. 
As a comparison to GRS~1915, IGR~J17091 was assumed to emit close to the Eddington luminosity, which implied either the BH has mass $<3\,M_{\rm \odot}$ or the system could be farther away than 20\,kpc \citep{Altamirano2011}. 
In previous studies, ten classes of variability have been identified by \cite{Court2017MNRAS.468.4748C} and \cite{Wang2024ApJ...963...14W}. Among them, two classes of the variability had been detected in the most recent outburst in 2022 \citep{Wang2024ApJ...963...14W}.
Moreover, this outburst has been detected in multi-wavelength from radio to X-ray with the telescopes including \atca, \neowise, \swift, \fermi, \nicer and \nustar.
Hence, the multi-wavelength dataset of IGR~J17091 makes it an ideal candidate for investigating the origin of the non-thermal emission in BHXRBs.

In this study, we present an analysis of the 2022 outburst of IGR~J17091 using the telescopes outlined above. Our specific objectives include studying the properties of the heartbeat-like variability and constraining the origin of the non-thermal emission during the heartbeat state. Specifically, we describe the observation and data reduction procedures in Section~\ref{sec:Methods}, present the results of X-ray timing and spectral analysis, along with a multi-wavelength SED analysis in Section~\ref{sec:Result}, discuss the implications of our results in Section~\ref{sec:Discussion}, and finally summarize in Section~\ref{sec:CONCLUSIONS}.

\begin{deluxetable*}{lccccccc}[!t]
\tablecaption{Summary of the \nustar observations of IGR~J17091.}
\label{tab:NuSTAR_obs}
\tablewidth{0pt}
\tablehead{
\colhead{ObsID} & \colhead{Date} & \colhead{Start Time} & \colhead{Exp. A} & \colhead{Exp. B} & \colhead{Count Rate A} & \colhead{Count Rate B}\\[-5pt]
 & & (MJD) & (ks)  & (ks)  & (ct\,s$^{-1}$) & (ct\,s$^{-1}$)} 
\startdata
80702315002 & 2022-03-23 & 59661.1693 & 14.8 & 15.0 & $84.7$ & $77.7$  \\
80702315004 (NV-1) & 2022-03-26 & 59664.5889 & 16.5 & 16.8 & $71.8$ & $66.2$  \\
80702315006 (NV-2) & 2022-03-29 & 59668.0149 & 11.9 & 11.3 & $95.1$ & $87.8$  \\
80802321002 & 2022-04-21 & 59690.2104 & 17.6 & 17.8 & $75.4$ & $69.3$  \\
80802321003 (NV-3) & 2022-06-16 & 59746.4750 & 16.1 & 16.8 & $69.6$ & $63.8$  \\
80802321005 & 2022-07-31 & 59791.8694 & 15.9 & 16.1 & $59.3$ & $54.0$  \\
80801324002 & 2022-08-22 & 59813.0544 & 27.6 & 27.8 & $38.8$ & $35.9$  \\
80801324004 (NV-4) & 2022-08-25 & 59816.6805 & 75.8 & 78.6 & $34.5$ & $31.9$  \\
\enddata
\end{deluxetable*}

\begin{figure*}[!t]
 \centering
   \includegraphics[width=\linewidth]{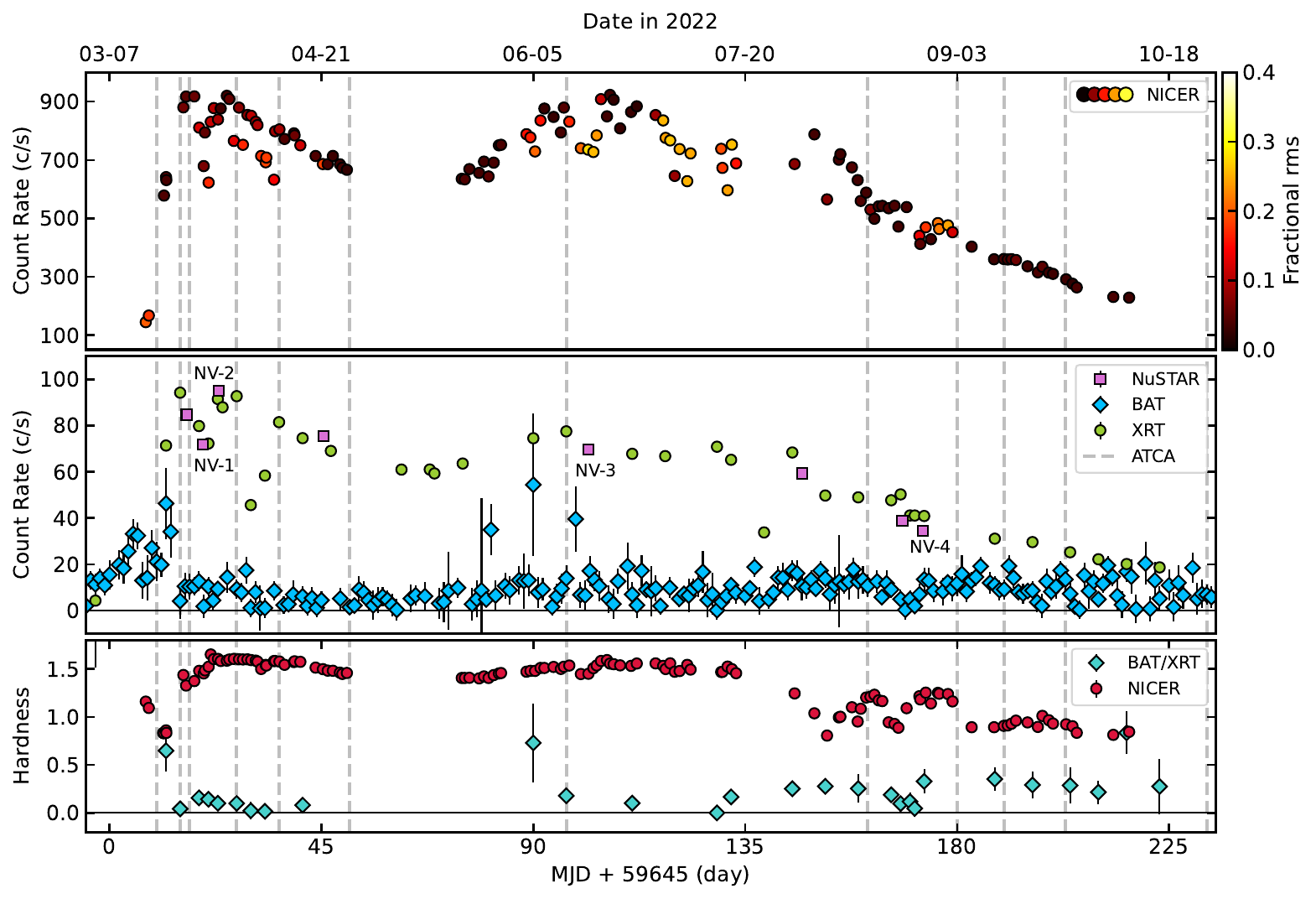}
   \caption{\textit{Top panel}: The lightcurve of \nicer during the 2022 outburst. The color bar represents the fractional rms \citep{Vaughan2003} in the 1--10\,keV band. \textit{Middle panel}: The lightcurves of \nustar (magenta), \swift-XRT (green), and \swift-BAT (blue). The dashed vertical lines denote the observation times with ATCA. \textit{Bottom panel}: The red circles and the cyan diamond represent the hardness ratio of \nicer count rates in the 2--10\,keV and the 1--2\,keV, and with \swift in the BAT (15--50\,keV) and XRT (0.3--10\,keV), respectively.}
   \label{fig:Swift_lc.png}
\end{figure*}

\begin{figure*}[!t]
   \centering
   \includegraphics[width=1\linewidth]{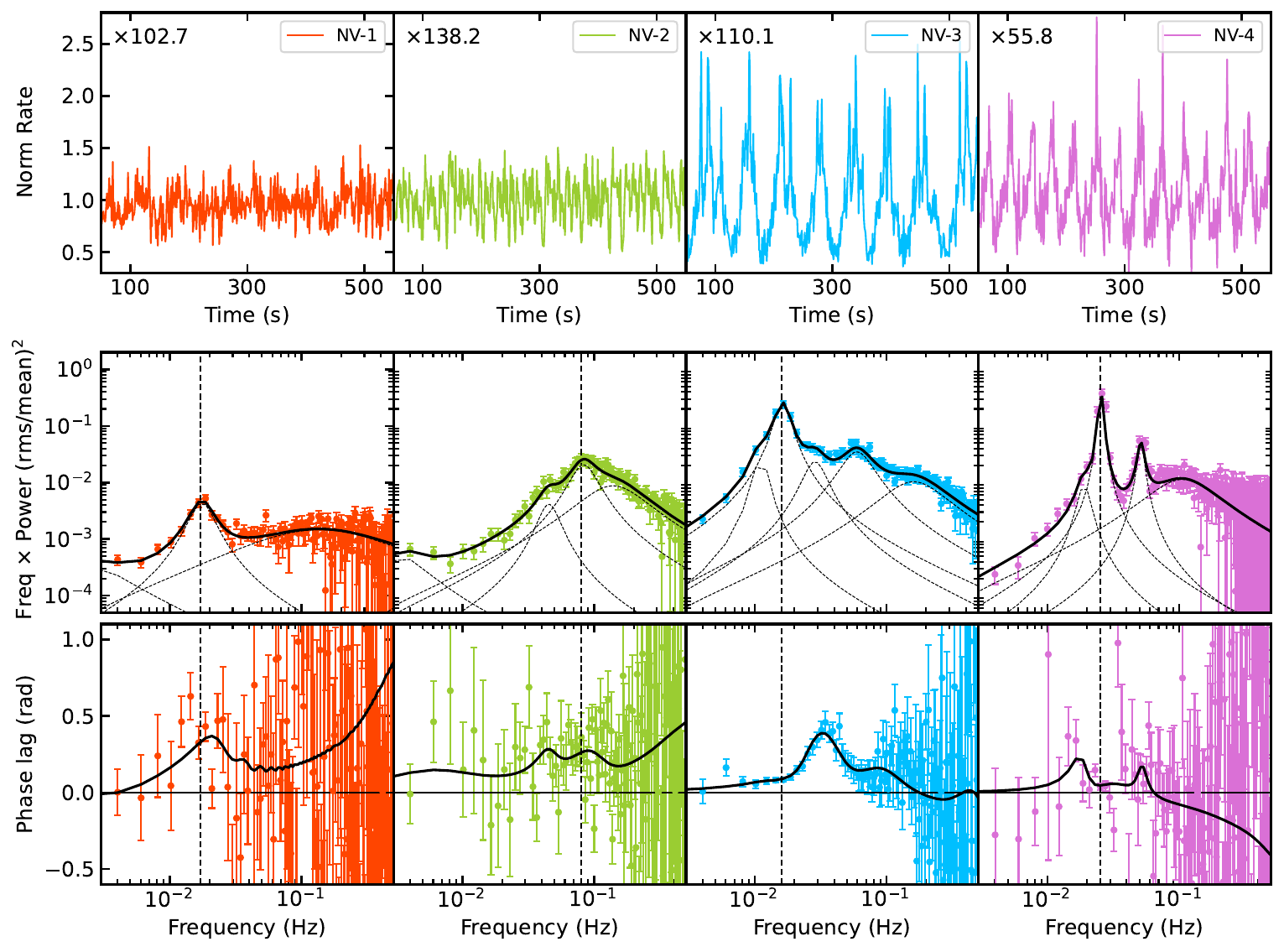}
   \caption{\textit{Top panel}: Normalized lightcurves of the four \nustar observations in FPMA (NV-1 in red, NV-2 in green, NV-3 in blue, and NV-4 in magenta). The normalization factor of count rate is shown in the top-left corner. \textit{Middle panel}: The co-spectrum between FPMA and FPMB in the energy band 3--50\,keV. The black solid curve denotes the overall best fitting model, and each dashed curve corresponds to an individual Lorentz component. The vertical dashed line marks the centroid frequency $\nu_\mathrm{c}$ of the heartbeat-like variability. \textit{Bottom panel}: The phase lag spectrum between 3--4\,keV (reference band) and 7--10\,keV. The black solid curve indicates the best fitting result of the constant time lag model \citep{Mariano2024MNRAS.527.9405M}.}
   \label{fig:lc_pds_lag}
\end{figure*}

\section{Observations and Data reduction} \label{sec:Methods}

\subsection{\nustar}
During 2022, eight \nustar observations of IGR~J17091 had been conducted (PI: J. Wang and J. Garcia). We conducted data processing using the \nustar Data Analysis Software (NUSTARDAS) version 1.9.7 with the Calibration Database version v20230420. We used the command {\it nupipeline} to calibrate the data with the arguments \textit{saamode=strict}, \textit{tentacle=yes}, and \textit{statusexpr=(STATUS==b0000xxx00xxxx000)\&\& (SHIELD==0)}.
We generated the source spectra and lightcurves for FPMA and FPMB in the energy range 3--50\,keV, respectively, along with their corresponding response and ancillary response files, using the {\it nuproducts} task. The extraction of the source region is circled with a radius of 100\arcsec, where the center is set at the emission peak by the {\it centroid} task. The background region is chosen by placing a circular at the farthest corner of the image from the source center, with the same radius as the source region. For observation 80801324004, a portion of the region in the FPMA image is contaminated, though the contamination does not affect the source region. This contamination, however, could lead to an overestimation of the background. Since this issue did not appear in FPMB, we chose to use the background file from FPMB as the background for FPMA for spectral fitting.

\subsection{\nicer}

We obtained a dataset comprising 175 observations from the \nicer telescope, spanning from March 14 to October 12, 2022. Standard calibration procedures and screening using the {\it nicerl2} task were applied to each observation. Subsequently, the lightcurves and spectra were generated with {\it nicerl3-lc}, and {\it nicerl3-spect} in which the background was estimated by the 3C50 tool \citep{Remillard2022}. 
The processing pipeline used the \nicer CALDB Version xti20221001.

\subsection{\swift}
The \swift/BAT Daily lightcurves were directly downloaded from the \swift/BAT Hard X-ray Transient Monitor\footnote{\url{https://swift.gsfc.nasa.gov/results/transients/}} \citep{Krimm2013}. The \swift-XRT lightcurves were generated by the online tools at UK \swift Science Data Centre\footnote{\url{https://www.swift.ac.uk/user\_objects/}} \citep{Evans2007,Evans2009}. 

Regarding the \swift-UVOT observations, we combined their sky images and exposure maps using the {\it uvotimsum} task for each filter, and generated the count rates using the {\it uvotsource} task with the UVOT calibration 20220705. The source region was placed at the center of the X-ray emission region with a radius of 5\arcsec. We carefully avoided any nearby sources and selected three background regions, each with a radius of 10\arcsec. 
However, there is no significant source was detected, and we therefore set a 3-$\sigma$ photometry upper limit at the source region of IGR~J17091.

\subsection{Fermi}
Following the official users' guide\footnote{\url{ https://fermi.gsfc.nasa.gov/ssc/data/analysis/scitools/binned_likelihood_tutorial.html}}, we performed a binned analysis using {\sc Fermitools 2.2.0} and the Pass 8 data covering MJD~59731--59832.
We used the {\it gtlike} tool to conduct the spectral analysis, during which we fixed parameters to the 4FGL-DR4 catalog values, except for the normalization of sources within 3 degree from our target. In addition, since IGR~J17091 is not in the 4FGL-DR4 catalog, we included it manually assuming a powerlaw spectrum with a photon index of two. This resulted in a low TS value for IGR~J17091, suggesting that it was not detected significantly.
Then we estimated its 0.1--10\,GeV flux upper limit $F_{0.1-10\, \mathrm{GeV}} = 3.3\times10^{-11}\, \mathrm{erg/cm^2/s}$
using the {\it UpperLimits}\footnote{\url{https://fermi.gsfc.nasa.gov/ssc/data/analysis/scitools/upper_limits.html}} tool assuming a 3-$\sigma$ confidence level.

\subsection{\neowise}
We searched for the \neowise archive data from 2022 for the infrared detection in the location of IRG~J17091. 
The W1 and W2 images were downloaded from the NASA/IPAC Infrared Science Archive\footnote{\url{https://irsa.ipac.caltech.edu/Missions/wise.html}} \citep{NeowiseI2020}. Unfortunately, our target was not significantly detected by \neowise in single visits. Hence, we stacked the images from August 2022 as the source and the images from 2019 to 2021, when the target was in quiescent state, as the background. Both of the images were selected within a circular region with a radius of 10\arcsec.
Eventually, we derived the absorbed fluxes $F_{\rm w1} = 1.41\pm0.54$\,mJy and $F_{\rm w2} = 0.68\pm0.59$\,mJy.

Each X-ray spectral data were grouped using the {\it grppha} task from {\sc ftools} package \citep{Blackburn1995} to achieve a minimum of 30 counts per bin, as the requirements for $\chi^2$ statistics. In our SED analysis, we apply the \cite{Fitzpatrick1999PASP..111...63F} dust extinction model with $\rm E(B-V) = 3.48$ \citep{Planck2016A&A...596A.109P} from the infrared to UV band. We adopt the \texttt{tbabs} component to account for the interstellar absorption, using the \textit{abund wilm} command to set the abundance table \citep{Wilms2000} and \textit{xsect vern} command to set the photoelectric cross sections \citep{Verner1996}. We fix the equivalent hydrogen column density at $N_\mathrm{H} = \rm 1.537\times10^{22}\,cm^{-2}$, as measured by the \nicer observations in 2022 \citep{Wang2024ApJ...963...14W}.
Unless explicitly mentioned, the uncertainties for each fitting parameter in this work were calculated at 1-$\sigma$ confidence level.

\section{Result} \label{sec:Result}

\subsection{Timing Analysis} \label{sec:Timing_Analyize}

\subsubsection{Lightcurves}

Fig.~\ref{fig:Swift_lc.png} shows the lightcurves of IGR~J17091 in the studied period, as observed with \nicer, \nustar, and \swift-XRT/BAT. We calculated the fractional rms \citep{Vaughan2003} for the \nicer's daily light curve with a time bin of one second to provide a general evolution of the variability amplitude, which indicates an intermittent occurrence of variability. Additionally, we observed quasi-periodic variability from lightcurve in \nustar observations ObsIDs 80702315004 (hereafter NV-1), 80702315006 (hereafter NV-2), and 80802321003 (hereafter NV-3). While during the two-day observation, ObsID. 80801324004 (hereafter NV-4), the variability was intermittent. Therefore, we only studied the observation segments of NV-4 where the variability persists for longer than 2000\,seconds. In this work, we refer to the variability with a period of tens of seconds and time lag of seconds as `heartbeat-like' variability (see Sect.~\ref{sec:Power_spectrum} for more details).

\subsubsection{Power spectrum and Cross spectrum} \label{sec:Power_spectrum}

\begin{deluxetable*}{lccccc}[]
\tablecaption{Properties of the heartbeat-like variability derived from the cross spectrum.}
\label{tab:PS_result}
\tablewidth{0pt}
\tablehead{
\colhead{ObsID} & \colhead{Class} & \colhead{$\nu_c$}  & \colhead{$\Delta \nu$} & \colhead{rms} & \colhead{Time lag$^a$} \\[-5pt]
& & (mHz)  & (mHz) & (\%) & (s)}
\startdata
80702315004 (NV-1) & -- &  $17.1 \pm 0.3$ & $7.1 \pm 1.1$ & $6.6 \pm 0.3$  & $3.7\pm0.6$\\
80702315006 (NV-2) &  V &  $79.8 \pm 2.0$ & $37.0 \pm 7.9$ & $15.7 \pm 1.8$ & $0.6\pm0.1$ \\
80802321003 (NV-3) &  X &  $16.0 \pm 0.2$ & $4.1 \pm 0.4$ & $42.2 \pm 1.7$ & $0.9\pm0.1$\\
80801324004 (NV-4) & IV &  $25.3 \pm 0.2$ & $1.1 \pm 0.2$ & $25.9 \pm 1.7$ & $0.3\pm0.1$\\ 
\hline
80802321003 (NV-3B) & -- &  $28.0 \pm 0.5$ & $11.8 \pm 1.3$ & $15.3 \pm 1.3$ & $2.9\pm0.3$\\
\enddata
\tablecomments{
$^a$: The time lag is calculated between 3--4\,keV and 7--10\,keV. A positive value means that high-energy photons arrive after low-energy ones.
}
\end{deluxetable*}

\begin{figure}[!t]
 \centering
   \includegraphics[width=1\linewidth]{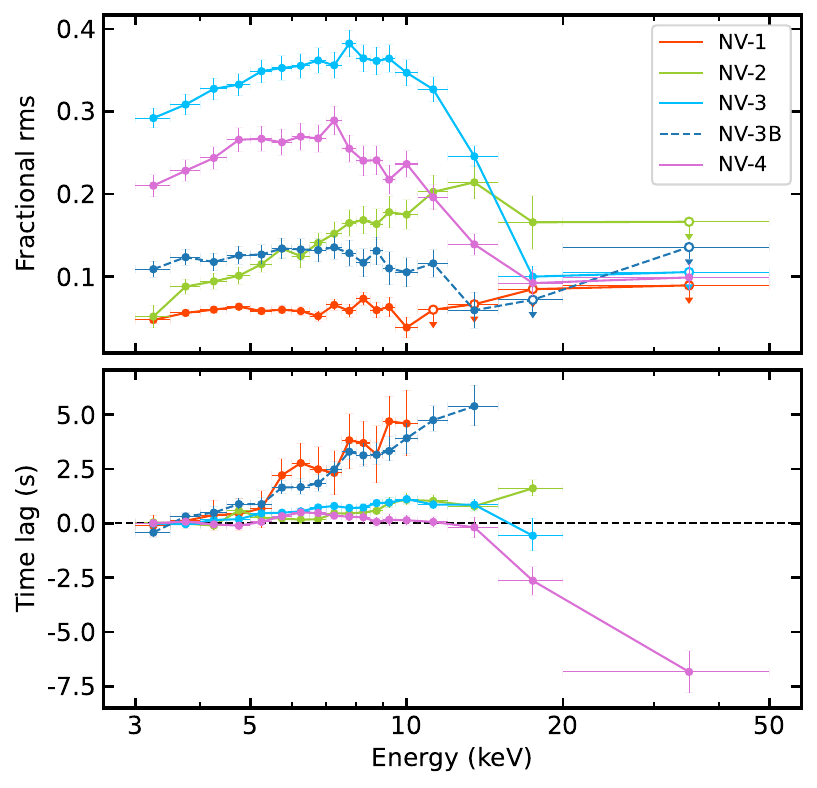}
   \caption{\textit{Upper panel}: The rms spectra of \nustar data. The solid curve is the fundamental Lorentzian component, while the dashed, dark blue curve is the NV-3B. The open circles mark the 3-$\sigma$ upper limit, where the Lorentz component is not significantly required. \textit{Lower panel}: The lag spectra of \nustar data. The inset plot in the lower-left corner provides a closer view along both the x- and y-axes.}
   \label{fig:rms_spe.png}
\end{figure}

In our timing analysis, we used the {\it AveragedPowerspectrum} and {\it AveragedCrossspectrum} packages in {\it Stingray} \citep{Huppenkothen2019-a,Huppenkothen2019-b} to generate the average power/cross spectra (PS/CS) from the lightcurves. We used the real part of the CS as the co-spectrum \citep{Bachetti2015}.
We applied logarithmic rebin with a factor of 0.01 and adopted the normalization with the fractional rms method \citep{Belloni1990}. For the \nustar data, the co-spectrum was generated over an energy range of 3--50\,keV between FPMA and FPMB, with a time step of $\delta t = 0.1$\,seconds and a segment length of $T = 500$\,seconds.
We further averaged the co-spectrum from each segment when there was no obvious shift in the frequencies of the main features of the co-spectrum. The shift confined within the full width at half maximum (FWHM), denoted as $\Delta \nu$, is considered acceptable.
To describe the co-spectrum, we used the {\it ftflx2xsp}\footnote{\url{https://heasarc.gsfc.nasa.gov/lheasoft/ftools/headas/ftflx2xsp.html}} task in the {\sc ftools} package \citep{Blackburn1995} to convert each co-spectrum data point into spectra and respose formats for fitting in {\sc xspec} \citep{arnaud1996}.
We fitted the co-spectrum by iteratively adding Lorentzian components until no prominent features remained in the residuals and the $\chi^2$ value was not significantly further improved. 
However, we found that NV-4 exhibits a significant shift in $\nu_c$ from 16\,mHz to 27\,mHz during the observation. Therefore, we included only the data for which the co-spectrum met the above criteria with no obvious shift in $\nu_c$, for the subsequent timing analysis.
The Lorentzian component with the strongest power at $\nu_c$ was identified as the fundamental component of heartbeat-like variability.
The fractional rms of the Lorentzian component was calculated by taking the square root of the Lorentz normalization factor, and a correction was applied to obtain the intrinsic rms using Eq.~(5) from  \cite{Bachetti2015}.
The middle panels in Fig.~\ref{fig:lc_pds_lag} display the Lorentz fitting results of the co-spectrum in the 3--50\,keV. The properties of the fundamental Lorentzian component are shown in Table~\ref{tab:PS_result}. We also generated the PS from the \nicer observations, where we noted that the heartbeat-like variability appears when the fractional rms is large, as shown in Fig.~\ref{fig:Swift_lc.png}.

According to the definition of the variability classes in GRS~1915 and IGR~J17091 by \cite{Belloni2000,Altamirano2011,Court2017MNRAS.468.4748C}, the heartbeat-like variability present in NV-2 and NV-4 corresponds to Class V and IV (or Class $\rho$), respectively. In addition, NV-3 is categorized as a new type of heartbeat-like variability, i.e., class X \citep{Wang2024ApJ...963...14W}. The amplitude of the variability in NV-1 is too weak to be further identified.

To obtain the rms spectra for each observation, we generated lightcurves in smaller energy bands from the \nustar event file and constructed their co-spectrum between FPMA and FPMB. We linked $\nu_c$ and $\Delta \nu$ of each Lorentzian component to the values derived in the full band co-spectrum, but let their Lorentzian normalization free to vary. 
At the energy band where the Lorentzian component was not significantly required, we provided a 3-$\sigma$ upper limit for its fractional rms. Regarding the lag spectra, we generated the CS/PS in the selected energy bands, where the Lorentzian component was significantly required, and used the 3--4\,keV band as the reference band.
We applied the constant time-lag model adopted from \cite{Mariano2024MNRAS.527.9405M} to jointly fit the PS, and the real and imaginary parts of the CS to determine the time lags.

We show the evolution of rms and time lag with energy in Fig.~\ref{fig:rms_spe.png}. The rms of NV-2, NV-3, and NV-4 share a similar trend: the rms increases with energy and then decreases after reaching an inflection energy, although the inflection energy differs among them. For NV-1, the rms spectrum remains nearly constant within the 3--10\,keV band. The lag spectra in all observations show a positive (hard) lag of several seconds below 15\,keV, while for NV-3 and NV-4, they change sign from positive to negative above 15\,keV.
Additionally, we detected a hump at $\sim$28\,mHz in the lag spectrum of NV-3, which could be explained by a Lorentzian component at $\nu_c = 28$\,mHz in the co-spectrum of NV-3. To distinguish it from the fundamental heartbeat-like component, we define this component as NV-3B. This component exhibits different evolutionary patterns in both rms and lag spectra from NV-3, but it shares similarities with the variability in NV-1.

Overall, the heartbeat-like variability in the hard energy band (above 20\,keV) has been observed to be either non-existent (for NV-1 to NV-3) or very weakly detected (for NV-4) compared to the soft energy band (below 20\,keV). This possibly suggests an inefficient propagation between soft and hard photons. 

\subsection{Spectral Analysis} \label{sec:Spectral_ana}
In the following, we conduct an analysis using average and phase-resolved spectra to further investigate the origin of the non-thermal X-ray emission.
\subsubsection{Average spectra}

\begin{figure}[!t]
 \centering
   \includegraphics[width=1\linewidth]{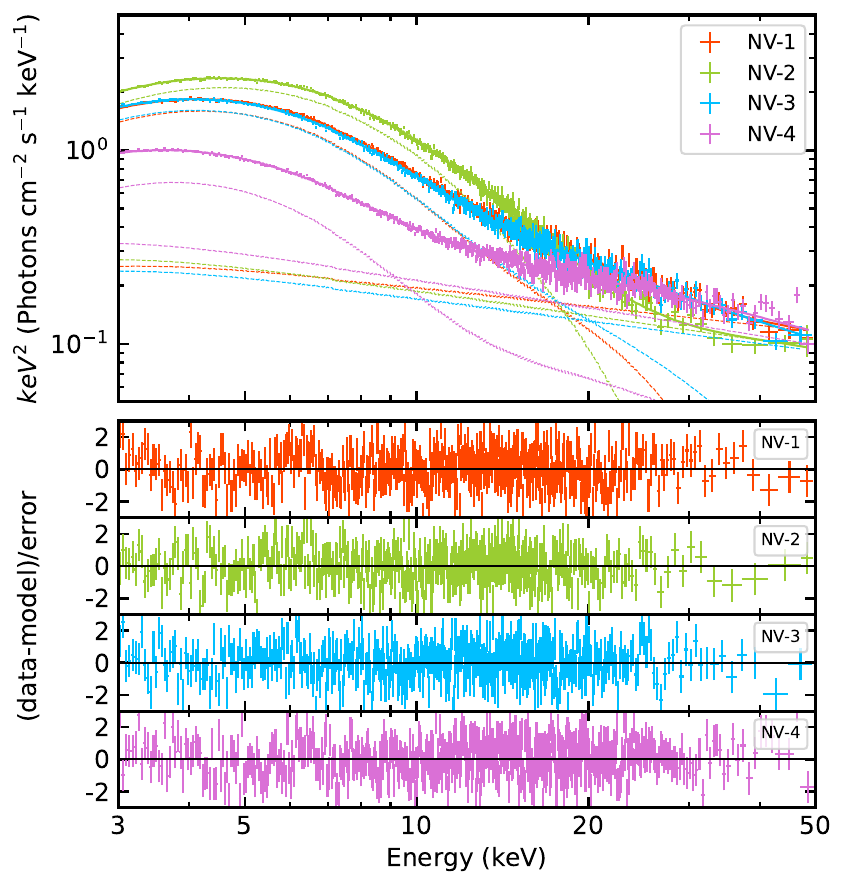}
   \caption{The unfolded \nustar spectra for FPMA and the corresponding residuals. The colors are defined the same as in Fig.~\ref{fig:lc_pds_lag}. The spectra of NV-1 and NV-3 overlap on the plot.}
   \label{fig:ave_spe.png}
\end{figure}

\begin{figure*}[!t]
 \centering
   \includegraphics[width=1\linewidth]{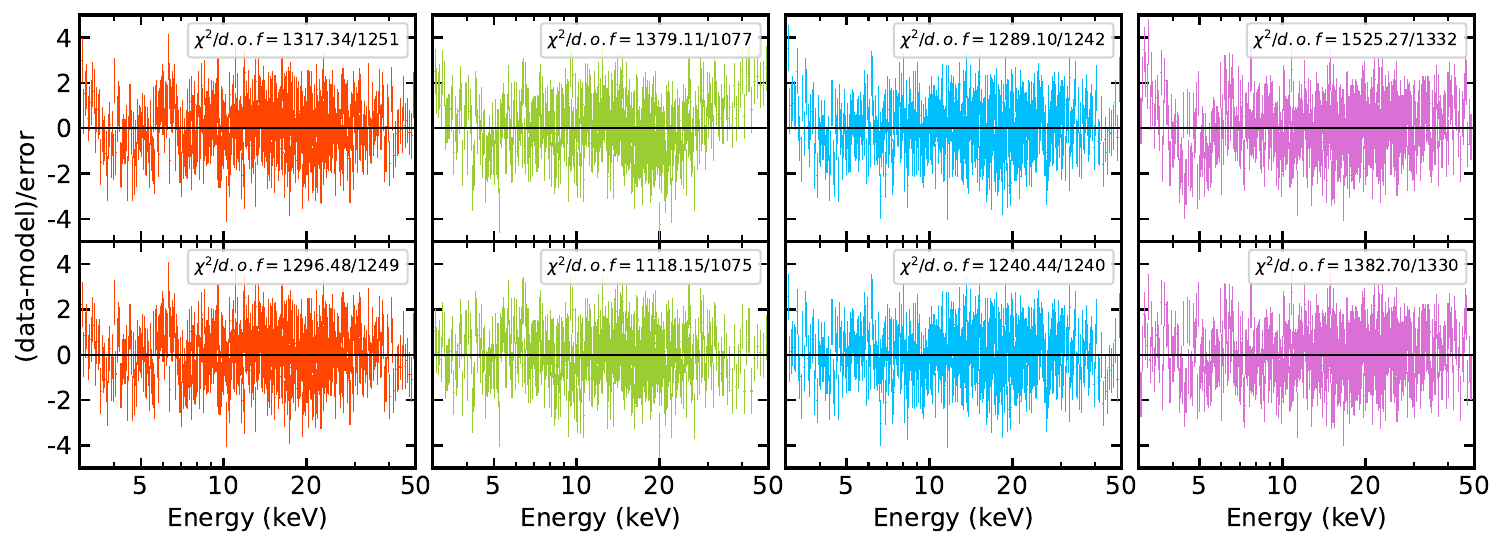}
   \caption{\textit{Top panel}: Residuals in FPMA from \nustar spectral fitting using the model: \texttt{constant} $\times$ \texttt{tbabs} $\times$ (\texttt{thcomp} $\times$ \texttt{diskbb}). The panels from left to right represent NV-1, NV-2, NV-3, and NV-4. The top-right value in each panel denotes the reduced $\chi^2$ for each fit. \textit{Bottom panel}: Residuals in FPMA from \nustar spectral fitting using the model: \texttt{constant} $\times$ \texttt{tbabs} $\times$ (\texttt{thcomp} $\times$ \texttt{diskbb} + \texttt{powerlaw}). }
   \label{fig:model_res.png}
\end{figure*}

The spectra were analyzed using {\sc xspec} version 12.12.1 \citep{arnaud1996}. We conducted spectral fitting for IGR~J17091 using \nustar data in the energy range of 3--50\,keV. We used an absorbed multi-color blackbody component \texttt{diskbb} \citep{Mitsuda1984} convolved by a \texttt{thcomp} model \citep{Zdziarski2020MNRAS.492.5234Z} to describe the disk emission and the inverse Compton scattering of the disk photons. We extended the energy range using the command {\it energies 0.01 1000.0 1000 log}. 
A multiplicative \texttt{constant} parameter was used to account for cross-normalization of the FPMA and FPMB instruments, where we fixed this to 1 for the FPMA and left it free for FPMB. 
The model is \texttt{constant} $\times$ \texttt{tbabs} $\times$ (\texttt{thcomp}$\times$\texttt{diskbb}). 
However, despite the reduced $\chi^2$ values ranging from 1.04 to 1.28, which appear marginally acceptable, there are still noticeable residuals in either the soft or the hard bands, as illustrated in the upper panels of Fig.~\ref{fig:model_res.png}.
To improve the fit, we incorporated a \texttt{powerlaw} component, which reduced the $\chi^2$ values to a range of 1.00 to 1.04 (see the lower panel of Fig.~\ref{fig:model_res.png}).
We further ran F-test to examine the significance of this \texttt{powerlaw} component, and obtained a p-value less than $10^{-5}$. This confirms that the inclusion of the additional \texttt{powerlaw} component is indeed necessary. 
The overall model is \texttt{tbabs} $\times$ (\texttt{thcomp} $\times$ \texttt{diskbb} + \texttt{powerlaw}). 

As the \texttt{thcomp} component is convolved with the \texttt{diskbb} component, it represents an inverse Comptonization process occurring in the corona. To investigate whether the \texttt{powerlaw} component is associated with the variability, we conducted phase-resolved spectroscopy.

\begin{figure*}[!th]
 \centering
   \includegraphics[width=0.495\linewidth]{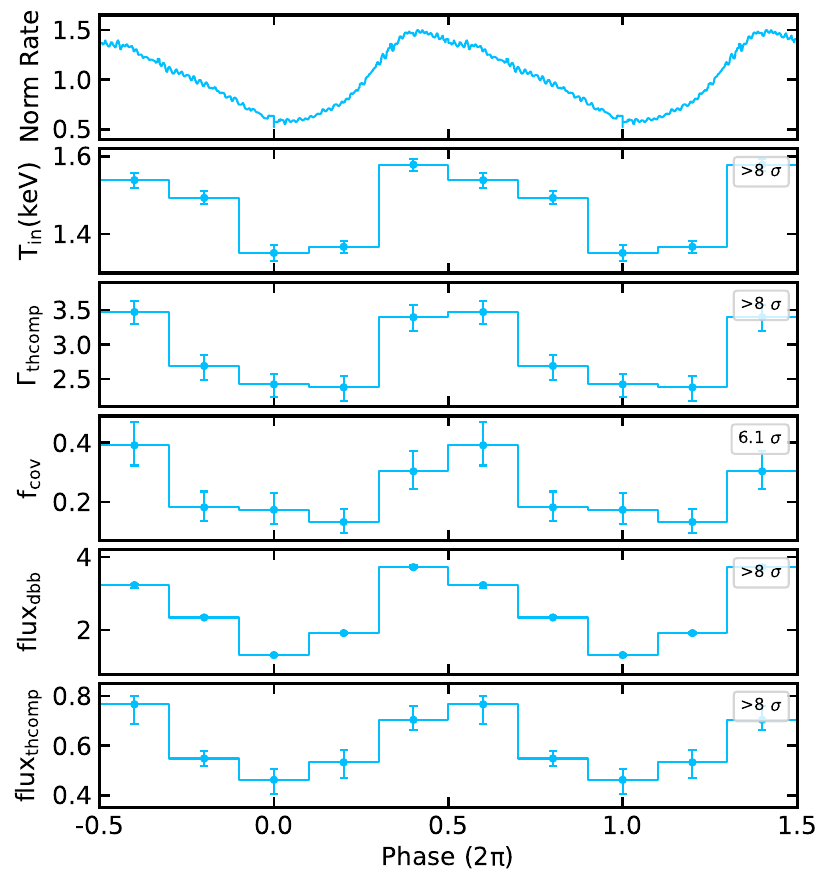}
   \includegraphics[width=0.495\linewidth]{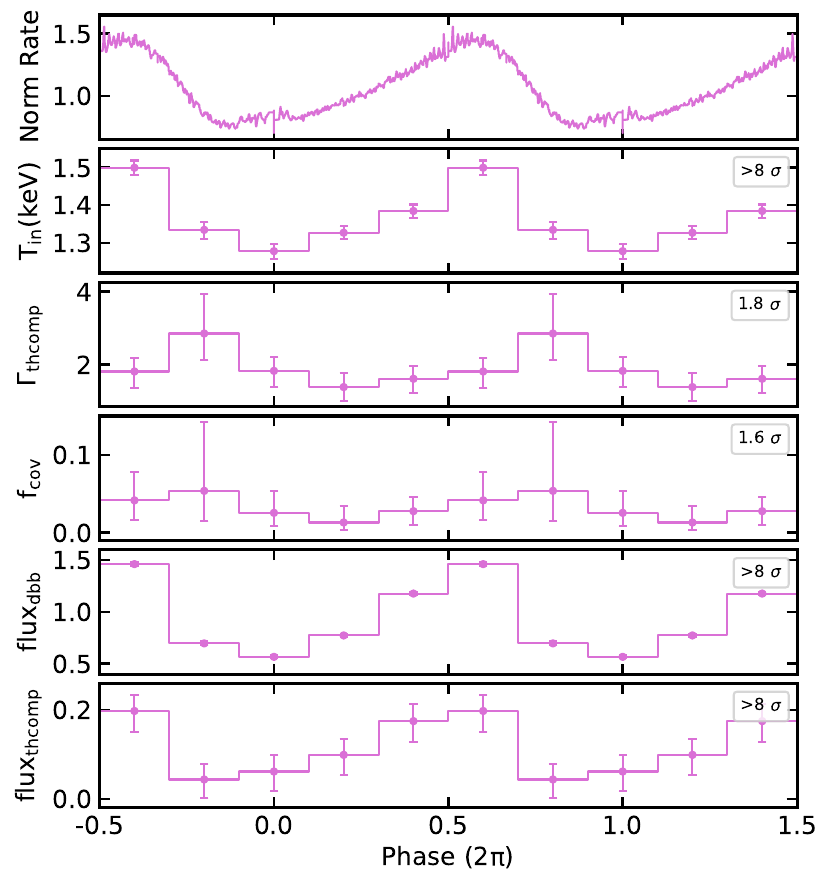}
   \caption{Best-fitting parameters from phase-resolved spectra of NV-3 and NV-4 using the model \texttt{constant} $\times$ \texttt{tbabs} $\times$ (\texttt{thcomp} $\times$ \texttt{diskbb} + \texttt{powerlaw}). The panels from top to bottom show: the normalized count rate, the disk temperature, the photon index of \texttt{thcomp}, the covering fraction, the unabsorbed flux of \texttt{diskbb}, and the unabsorbed flux of \texttt{thcomp}. The legend in the top-right corner shows the significance of parameter variations, assessed by allowing the parameters to vary freely and linking them across all phases using the F-test.}
   \label{fig:616_phase_result}
\end{figure*}

\begin{deluxetable*}{lcccccc}[!ht]
\tablecaption{Best-fitting parameters of the \nustar spectra of IGR~J17091.}
\tablewidth{0pt}
\tablehead{
\colhead{Parameters} & \multicolumn{4}{c}{\nustar ObsID}\\
 &  80702315004 & 80702315006  &  80802321003 & 80801324004 \\
 &  NV-1 & NV-2  &  NV-3$^a$ & NV-4$^a$
}
\startdata
  & & \\[-8pt]
$C^{b}_{\rm FPMB}$ & $0.992\pm0.001$ & $0.988\pm0.001$ & $0.981\pm0.001$ & $0.987\pm0.001$ \\[4pt]
\hline & & \\[-8pt]
$N_{\rm H}^c$\,$({\rm cm^{-2}})$ & \multicolumn{4}{c}{$1.537\times10^{22}$}\\[4pt]
\hline & & \\[-8pt]
$T_{\rm in}$\,(keV) & $1.57^{+0.04}_{-0.03}$ & $1.70^{+0.03}_{-0.04}$ & $1.47\pm0.01$ & $1.38\pm0.02$ \\[4pt]
$N_{\rm dbb}$ & $41.2^{+3.8}_{-4.1}$ & $37.8^{+3.1}_{-2.3}$ & $52.6^{+2.4}_{-2.4}$ & $28.5^{+2.0}_{-1.7}$ \\[4pt]
\hline & & \\[-8pt]
$\Gamma_{\rm thcomp}$ & $1.7^{+0.5}_{-0.7^{d}}$ & $1.3^{+0.8}_{-0.3^{d}}$ & $2.9^{+0.1}_{-0.2}$ & $1.7^{+0.3}_{-0.4}$ \\[4pt]
$kT_{\rm e}$\,(keV) & $3.8^{+0.9}_{-0.6}$ & $2.5^{+0.3}_{-0.1}$ & $9.4^{+6.0}_{-2.6}$ & $4.5^{+1.2}_{-0.6}$ \\[4pt]
$f_{\rm cov}$ & $0.05^{+0.06}_{-0.04}$ & $0.03^{+0.14}_{-0.02}$ & $0.3\pm0.1$ & $0.03\pm0.02$ \\[4pt]
\hline & & \\[-8pt]
$\Gamma_{\rm pl}$ & $2.3\pm0.1$ & $2.4\pm0.1$ & $2.4\pm0.1$ & $2.46\pm0.01$ \\[4pt]
$N_{\rm pl}$ & $0.4\pm0.1$ & $0.5\pm0.1$ & $0.4\pm0.1$ & $0.73^{+0.04}_{-0.05}$ \\[4pt]
\hline & & \\[-8pt]
$F_{\rm dbb}^{e}$& $2.55^{+0.06}_{-0.03}$ & $3.5\pm0.1$ & $2.40^{+0.01}_{-0.03}$ & $0.91\pm0.01$\\[4pt]
$F_{\rm thcomp}^{e}$& $0.4\pm0.1$ & $0.6\pm0.1$ & $0.6\pm0.1$ & $0.13^{+0.04}_{-0.03}$\\[4pt]
$F_{\rm pl}^{e}$ & $0.86\pm0.04$ & $0.8\pm0.1$ & $0.7^{+0.1}_{-0.2}$ & $1.11\pm0.04$ \\[4pt]
\hline & & \\[-8pt]
$\chi^2$/dof  &  $1296.5/1249$ (1.04) & $1118.2/1075$ (1.04) &  $5126.6/5025$ $(1.02)$ & $5948.6/5832$ $(1.02)$ \\[4pt]
\enddata
\tablecomments{
$^a$: The results for NV-3 and NV-4 are obtained through the joint fitting of the average and phase-resolved spectra;\\
$^b$: The constant parameter for FPMA is fixed at 1;\\
$^c$: $N_{\rm H}$ is adopted from \cite{Wang2024ApJ...963...14W};\\ 
$^d$: $\Gamma_{\rm th}$ pegs at its hard limit of 1.001;\\
$^e$: The unabsorbed flux is in units of $10^{-9}$\,erg\,cm$^{-2}$\,s$^{-1}$.\\
}
\label{tab:average_spe_fit}
\end{deluxetable*}

\subsubsection{Phase-resolved spectra}\label{sec:pr_spe}
Due to the low rms amplitude and the high $\Delta \nu$ (see Table~\ref{tab:PS_result}) of the variability in NV-1 and NV-2, it is difficult to accurately determine the variability profile from their lightcurves and hence the phase. Thus, we only did phase-resolved spectroscopy for NV-3 and NV-4. To eliminate the noise contribution to the heartbeat-like variability, we applied an optimal filtering algorithm to the fundamental Lorentz component of NV-3 and NV-4 (see Table~\ref{tab:PS_result}) in the 3--50\,keV lightcurve, as described by \cite{vandenEijnden2016MNRAS.458.3655V}. 
We then used the {\it find\_peaks} package from the {\it scipy} \citep{Virtanen2020} library to identify dips within each sine-like pattern present in the filtered lightcurve.
The nearest two dips were designated as phases $\phi=0$ and $\phi=2\pi$. We finally folded the original lightcurve and assigned five phases accordingly. 

We fitted the phase-resolved spectra with the same model as used for fitting the average spectra, i.e. \texttt{constant} $\times$ \texttt{tbabs} $\times$ (\texttt{thcomp} $\times$ \texttt{diskbb} + \texttt{powerlaw}). 
Due to the limitation of the statistics, we jointly fitted the average and the phase-resolved spectra for NV-3 and NV-4 to improve the constraints on the parameters. 
To examine the variation of each parameter across different phases, we linked each parameter and quantitatively assessed their respective changes by $\chi^2$ values. We find that linking the \texttt{powerlaw} component and the electron temperature ($kT_e$) in \texttt{thcomp} across different phases do not significantly affect the $\chi^2$ values. 
Hence, we linked the \texttt{powerlaw} component and $kT_e$ among each phase.
This approach resulted in a slight increase in the $\chi^2$, i.e. $\Delta \chi^2$ = 25.0 for 15 additional degrees of freedom (dof) in NV-3 and $\Delta \chi^2$ = 31.5 for 15 additional dof in NV-4.
Thus, these results indicate that the \texttt{powerlaw} component and $kT_e$ in \texttt{thcomp} do not exhibit significant variations across different phases.
We also calculated the flux for each component from the fits to both the average and the phase-resolved spectra with the command {\it cflux}.
 
The best-fitting parameters of the average spectrum for each observation are presented in Table~\ref{tab:average_spe_fit}, and the corresponding spectra, as well as the residuals, are shown in Fig.~\ref{fig:ave_spe.png}. We observe a high disk temperature ($T_{in} > 1.3$\,keV) and a low electron temperature of the corona ($kT_e < 10$\,keV) during the heartbeat-like variability.
We show the best-fitting parameters derived from the phase-resolved spectroscopy as a function of phase in Figs.~\ref{fig:616_phase_result}. 
The folded lightcurve of NV-3 presents a profile characterized by a rapid rise and slow decay, whereas NV-4 displays a slow rise and a rapid decay. In both NV-3 and NV-4, the evolution of the disk temperature aligns with the lightcurve profile. In addition, the flux difference between the peak and dip phases in \texttt{diskbb} is about five times greater than in \texttt{thcomp}, implying that the heartbeat-like variability is dominated by the accretion disk.

\subsubsection{Multi-wavelength SED} \label{sec:SED}
The results above suggest that at least part of the non-thermal emission, i.e. the \texttt{powerlaw} component, may not be associated with the thermally originated heartbeat-like variability, and hence not be attributed to the inverse Componization of the disk photons. To investigate whether this component could originate from jets, we conducted a broadband SED fitting spanning from radio to X-rays.

\atca observed IGR~J17091 several times in 2022 with the dates marked by the vertical dashed lines in Fig.~\ref{fig:Swift_lc.png}. Only two of them, June 12 (Projects: C3456) and September 3, 2022 (Projects: CX501), were quasi-simultaneously with our well-defined heartbeat-like variability in NV-3 and NV-4. 
According to the flux density measured by Russell et al. (in prep.), IGR~J17091 was not detected on June 12, with 3-$\sigma$ upper limits of 126\,$\mu$Jy at 5.5\,GHz and 106\,$\mu$Jy at 9\,GHz. On the other hand, IGR~J17091 was detected on September 3, with a flux density of $170 \pm 35$\,$\mu$Jy at 5.5\,GHz and $151 \pm 38$\,$\mu$Jy at 9\,GHz, respectively. The ATCA detection on September 3 was used in our SED fitting.

To extract the \texttt{powerlaw} component from the X-ray spectrum, we subtracted the contribution from the \texttt{thcomp} $\times$ \texttt{diskbb} component from the \nustar data and divided it with \texttt{tbabs}, using the best-fitting parameters from Table~\ref{tab:average_spe_fit}. 
Since we only fitted the continuous broadband spectrum, the \nustar data were further rebinned by a factor of 0.1 in logarithmic space. 
Moreover, due to the uncertainties in dust extinction and potential measurement bias in the infrared photometry, we excluded it from the SED analysis below. However, we used it together with the upper limits of the \swift-UVOT and \fermi data to evaluate the overall fit. We considered a leptonic jet model to fit and interpret the radio and X-ray data. For this, we used the open-source \textit{JetSeT}\footnote{\url{https://jetset.readthedocs.io/en/latest/}} framework \citep{Tramacere2009,Tramacere2011,Tramacere2020}, which includes the synchrotron and SSC processes.

In the adopted jet model, we considered a basic assumption of power-law energy distribution of relativistic electrons. This leads to eight parameters in the model: the minimum Lorentz factor, $\gamma_{\rm min}$, the minimum Lorentz factor, $\gamma_{\rm max}$, the spectral index of relativistic electrons distribution, $p$, the Doppler beaming factor, $\delta_D$, the magnetic field intensity $B$, the radius of the emitting blob, $R$, the total electron number density in the blob, $N_\mathrm{e}$, and the luminosity distance, $D$\footnote{The original parameter required by the \textit{JetSeT} model is the cosmological redshift. We converted it to the luminosity distance here.}.

The further setup for the model is as follows. 
Considering the previous measurements of the viewing angle, $\theta = 45^{\circ}.3\pm0^{\circ}.7$ \citep{Wang2018MNRAS.478.4837W} and $\theta = 24^{\circ}\pm4^{\circ}$ \citep{Wang2024ApJ...963...14W}, which suggests a Doppler beaming factor $\delta_D < 2.5$, we initially adopted the Doppler factor of $\delta_D = 1$. 
According to particle acceleration simulations \citep[e.g.][]{Sironi2011,Sironi2014}, we chose $\gamma_{\rm min} \leq 100$.
Regarding $\gamma_{\rm max}$, previous studies suggest that jets can be efficient at accelerating particles to energies above 10\,GeV in microquasars \citep[e.g.,][]{Bosch2009, Molina2019, Harvey2022}, and hence $\gamma_{\rm max}=10^5$ is adopted. This enables the SED to extend up to energies of $\approx$50\,GeV, within the \fermi range.
The luminosity distance of IGR~J17091 is still uncertain. Here, we adopted a luminosity distance of 13.7\,kpc, estimated from the radio–X-ray relationship (see Sect.~\ref{sec:RX_plane} for details).  
Ultimately, we are left with five free parameters for the fitting process: $\gamma_{\rm min}$, $p$, $B$, $R$, and $N_\mathrm{e}$. 

Moreover, we used the {\it Minuit ModelMinimizer} \citep{James1975} option in \textit{JetSeT} to provide initial values for the model parameters, and used the Monte Carlo Markov Chain (MCMC) ensemble sampler \citep{Foreman-Mackey2013} to archive a robust fit. This involves initializing 50 walkers to solve the maximum likelihood solution and exploring the parameter space with $10^4$ steps each. The parameter $B$, $R$, and $N_e$ were set in log-scale during the fitting.

\begin{figure}[!t]
   \centering
   \includegraphics[width=1\linewidth]{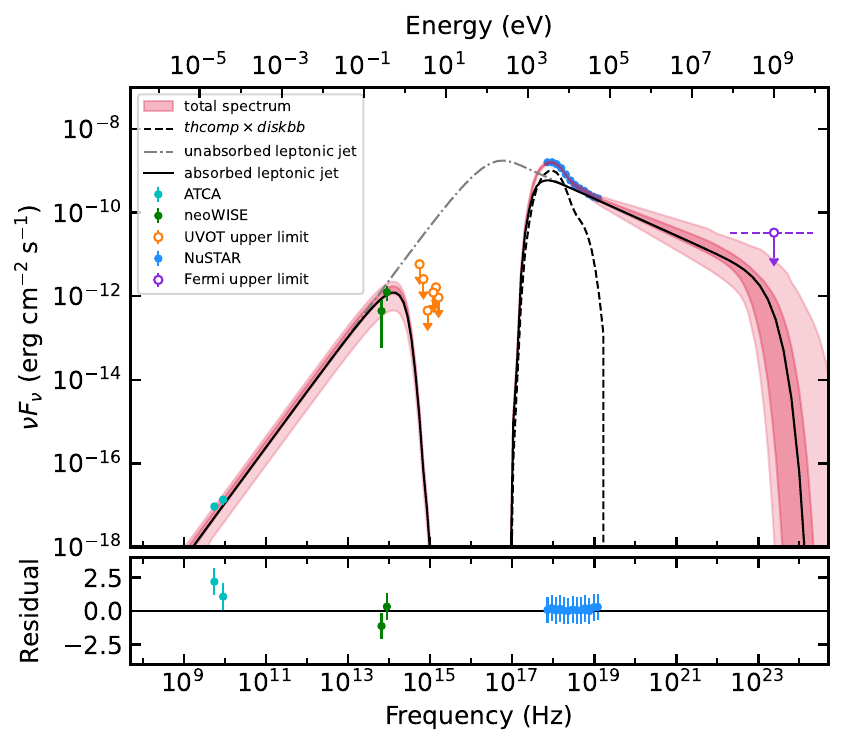}
   \caption{Top panel: Absorbed SED fitted with a synchrotron model (case (a) in Sect.~\ref{sec:SED}). The red region represents the total spectrum of the last 1,000 steps of MCMC walkers within the 1-$\sigma$ and 3-$\sigma$ confidence intervals. The solid line shows the best-fitting absorbed leptonic jet model, while the dashed line corresponds to the \texttt{thcomp} $\times$ \texttt{diskbb} model. The data points are from \atca (cyan), \neowise (green), UVOT (orange), \nustar (blue), and \fermi-LAT (purple). The horizontal purple dashed line on the \fermi-LAT data represents the instrument's energy range. The broad gap from $10^{15}$\,Hz to $10^{17}$\,Hz is due to high extinction, and the unabsorbed leptonic jet model is plotted as the grey dash-dot line. Bottom panel: Residuals of the best-fitting model.}
   \label{fig:sync.png}
\end{figure}

\begin{figure}[!ht]
   \centering
   \includegraphics[width=1\linewidth]{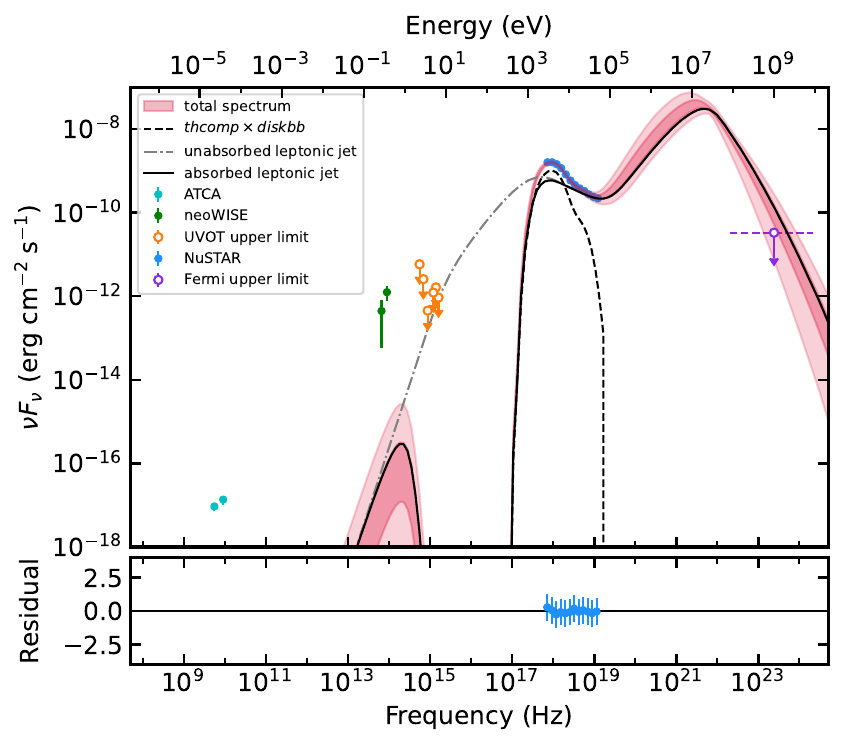}
   \includegraphics[width=1\linewidth]{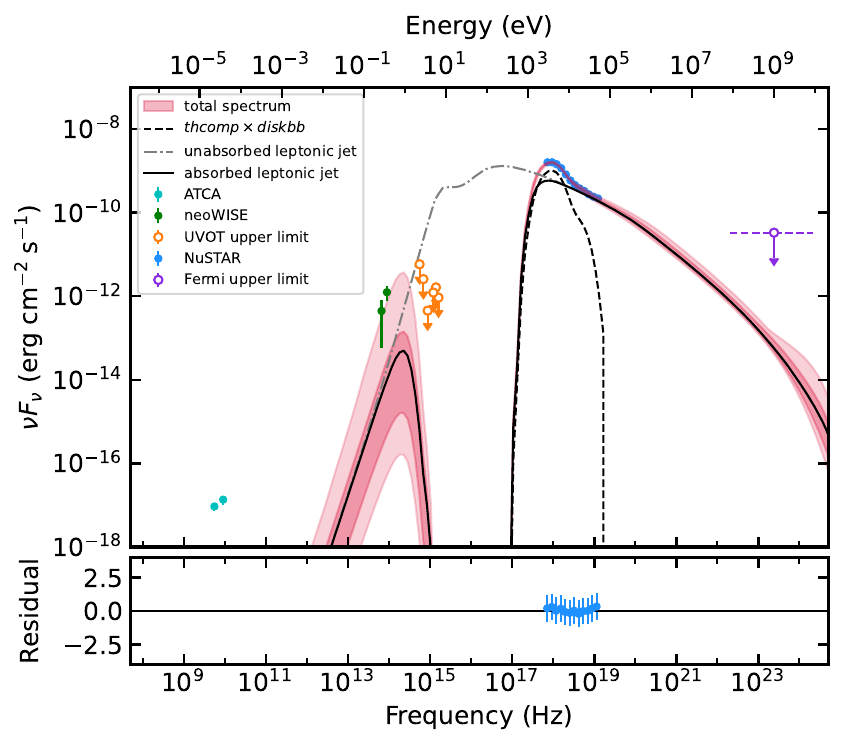}
   \caption{Absorbed SEDs and their residuals for cases (b) (top panel) and (c) (bottom panel) described in Sect.~\ref{sec:SED}. The colors are the same as in Fig.~\ref{fig:sync.png}.}
   \label{fig:sync_bc.png}
\end{figure}

Upon the fit, we obtain $p = 3.88 \pm 0.09$, consistent with the result shown in Table~\ref{tab:average_spe_fit}, where $\Gamma_{\text{po}} = (p + 1)/2$. If further assuming that both radio and X-ray emissions originate from the same synchrotron process (referred to as case (a) subsequently), we constrain $R > 10^{12}$\,cm and $N_e < 5$\,$\rm cm^{-3}$, though both parameters are correlated as $N_e \, R^3 \approx 1.2 \times 10^{36}$ (shown in the cornerplot in Fig.~\ref{fig:MCMC_c.png}). The values of $\gamma_\mathrm{min}$ and $B$ are also related as $B\approx4.5\times10^7 \left( \frac{10}{\gamma_\mathrm{min}} \right)^2 $\,G (also seen in Fig.~\ref{fig:MCMC_c.png}). As $1.1 <\gamma_\mathrm{min}< 100$, $B$ would be in the range of $10^5$--$10^9$\,G. 
We show the best-fitting SED in Fig.~\ref{fig:sync.png} and the parameter distributions in the left panel of Fig.~\ref{fig:MCMC_c.png}.

However, this assumption results in a very large magnetic energy, $E_B \sim B^2 \, R^3/6 > 3\times10^{46}$--$10^{54}$\,erg, which seems to be impractical in a BHXRB. This also leads to an extremely large ratio between the magnetic and electron energy densities, $U_B/U_e > 10^{10}$. Overall, we exclude the possibility that both the radio and X-ray emissions observed in IGR~J17091 originate from the same synchrotron process. 

We then consider two other cases: (b) the X-ray emission is produced by synchrotron processes in a compact spherical blob, whereas the radio emission comes from a more expanding region, and (c) the same as above but the X-ray emission is from SSC. We note that these two possibilities would require a multi-zone jet model \citep[e.g.,][]{Kaiser2006,Kantzas2021,Lucchini2022,Tramacere2022}, but for simplicity we neglect the fitting of the radio data which would constitute an upper limit to the compact blob emission, and focus only on the analysis of the high-energy emission within a one-zone approximation. 

Here, we assume an energy equipartition condition, $U_B = U_e$. In this case, $N_e$ is calculated as follows:
\begin{equation}
    N_e = \frac{B^2}{8\pi m_e c^2} \times \frac{(\gamma_\mathrm{max}^{1-p}-\gamma_\mathrm{min}^{1-p})/(1-p)}{(\gamma_\mathrm{max}^{2-p}-\gamma_\mathrm{min}^{2-p})/(2-p)}.
\end{equation}
The fitting allows for two sets of solutions, one with $\gamma_\mathrm{min} >50$ that corresponds to case (b), and one with $\gamma_\mathrm{min} < 10$ that corresponds to case (c).

In case (b), the X-ray emission is dominated by synchrotron, while SSC dominates above 100\,keV (see the upper panel of Fig.~\ref{fig:sync_bc.png}.). This fit then yields a gamma-ray flux of $F_{0.1-10\, \mathrm{GeV}} = 2.5_{-0.8}^{+0.7}\times10^{-9}$\,\flux, significantly exceeding the \fermi upper limit of $3.3\times10^{-11}$\,\flux. Therefore, we disfavor this case.

In case (c), the X-ray emission should originate from SSC (see the lower panel of Fig.~\ref{fig:sync_bc.png}). The obtained best-fitting parameters are $\gamma_\mathrm{min} < 6.4$, $p=4.3\pm0.2$, $B = (0.3$--$3.5)\times10^6$\,G, and $R = (0.1$--$2.5)\times10^8$\,cm, where the parameters $B $ and $R$ are actually correlated. Namely, the SSC flux is approximately $\propto B R^{-6/(p+5)}$ \citep{Shidatsu2011}, leading to $B \approx 9.3\times10^{10}\,R^{-0.65} $\,G (see the right panel of Fig.~\ref{fig:MCMC_c.png}). It is worth noting that variations in $D$ (11--23\,kpc) and $\delta_D$ (1--2.5) do not alter our conclusions.

\begin{figure}
   \centering
   \includegraphics[width=1\linewidth]{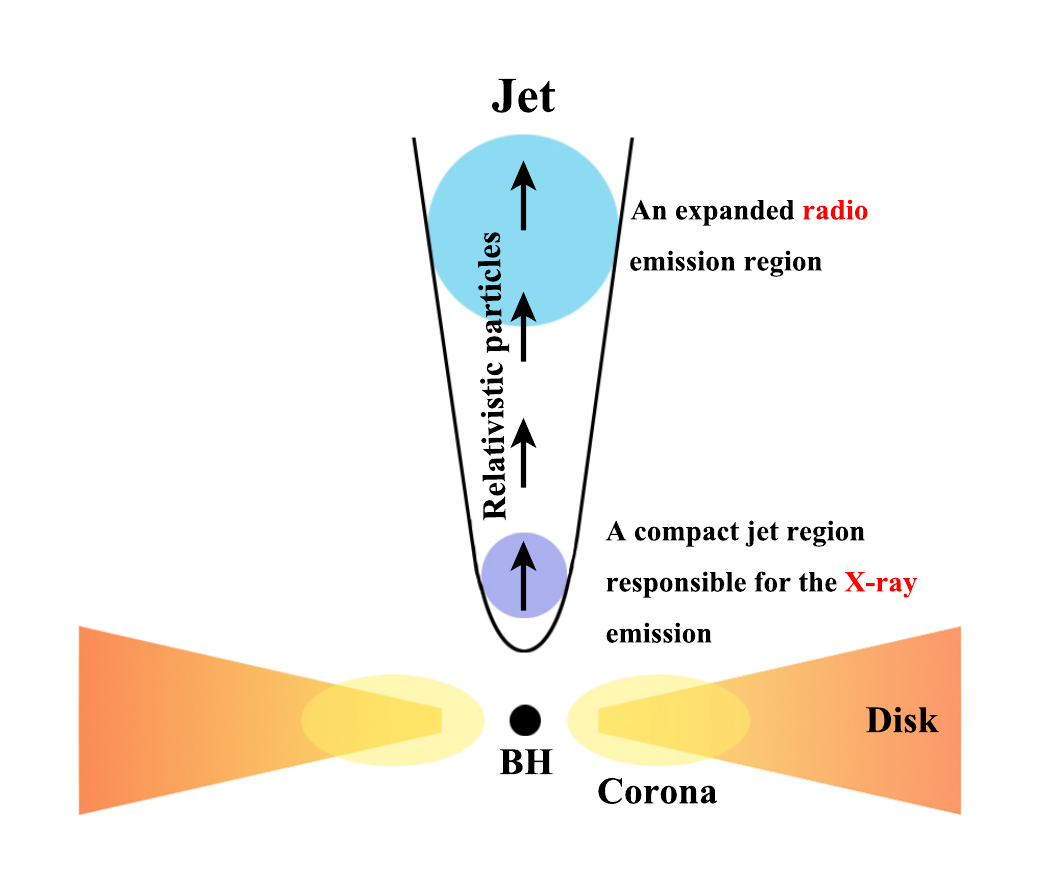}
   \caption{A schematic picture of the preferred case (c). The model consists of a multi-zone jet: around the base (the small purple region), the jet is optically thick at radio frequencies and the X-ray emission is produced via SSC; as the jet further expands, it becomes more transparent, eventually leading to the emergence of synchrotron radio emission (the large light-blue region).}
   \label{fig:sketch.png}
\end{figure}

Overall, our results favor case (c), where the X-ray emission in the jet originates from the SSC process within a compact blob with a strong magnetic field, while the radio emission arises from a more expanding region distinct from the X-ray emitting region. A simple schematic of case (c) is provided in Fig.~\ref{fig:sketch.png}.

\section{Discussion} \label{sec:Discussion}
In this study, we undertake a comprehensive timing and spectral analysis of the heartbeat-like variability observed in IGR~J17091 during the 2022 outburst. 
We employ this type of variability as an indirect method to differentiate the origins of non-thermal components emanating from the corona and jets. Our phase-resolved spectroscopy reveals the presence of a power-law component that remains uncorrelated with the heartbeat-like variability associated with the accretion disk. Here, we discuss the potential origins of this variability, the nature of the non-thermal components, and their broader implications.

\subsection{The origin of the heartbeat-like variability} \label{sec:Discussion_ori_HB}
As mentioned in Section~\ref{sec:Timing_Analyize}, at least three classes of heartbeat-like variability were detected in the four studied \nustar observations. These variabilities exhibit a similar evolutionary pattern in the rms spectrum, where the rms initially increases with energy and then decreases at higher energies. Although the inflection energy varies among observations, this value appears to correlate with the disk behavior; specifically, a higher disk temperature and a larger disk flux ratio correspond to a higher inflection energy (see Fig.~\ref{fig:rms_spe.png} and Table~\ref{tab:average_spe_fit}). In the standard disk blackbody model, where the disk temperature ($T_{\rm in}$) is proportional to the inner disk radius ($R_{\rm in}$) raised to the power of -3/4 \citep[e.g.,][]{Shakura1973,Makishima1986}, fluctuations in $R_{\rm in}$ lead to changes in $T_{\rm in}$. Consequently, this would result in a monotonic increase in the fractional rms with energy beyond $T_{\rm in}$. The observed decrease in fractional rms at higher energy bands suggests the presence of an additional component, which is more prominent in flux and acts to dilute the variability originating from the disk. 

Moreover, the detected time lag between 3--4\,keV and energies above 5\,keV is up to several seconds. Such a lag is at least two orders of magnitude larger than those observed for low-frequency QPOs \citep[e.g.,][]{Zhang2017,Zhang2020,Mariano2022,Nathan2022,Wang2022ApJ...930...18W} and other types of variabilities at similar frequencies \citep[e.g.,][]{Ma2021NatAs...5...94M,Liu2022}. Therefore, we can rule out the possibilities of geometry and intrinsic variability in the corona, light travel time between the disk and corona, and the inverse Compton scattering of disk photons in the corona \citep[e.g.,][]{Zdziarski1985,Miyamoto1988,Kara2019,Wang2022ApJ...930...18W}. 
However, the timescale of the lag is consistent with the order of the viscous timescale for matter transfer in the inner region of the disk \citep[e.g.,][]{Mir2016}. Combined with the correlation between the disk temperature and the count rate  (Fig.~\ref{fig:616_phase_result}), all the evidence supports the heartbeat-like variability originating from radiation pressure instabilities in the accretion disk \citep[e.g.,][]{Janiuk2000,Nayakshin2000,Done2007}. 

Regarding the new type variability, Class X, observed in NV-3, it actually exhibits a similar rms/lag evolution with that of Class IV in NV-4, but with higher rms below $\sim$15\,keV.
Its folded lightcurve presents a profile characterized by a rapid rise and slow decay, opposite to NV-4. 
Interestingly, we observed a significant signal (NV-3B) in its lag spectrum, corresponding to a much weaker Lorentzian component at $\nu_c = 28 \pm 0.5$\,mHz in the CS (see the light blue data points in Fig.~\ref{fig:lc_pds_lag}). 
Furthermore, the lag/rms spectrum of NV-3B evolves differently from that of its fundamental component, i.e., NV-3 (Fig.~\ref{fig:rms_spe.png}). All of these phenomena suggest that this additional variability may originate from a different emission region. However, further exploration is hindered by the nearby, dominant heartbeat-like variability in NV-3.

\subsection{The origins of the non-thermal emission in XRBs}

As discussed in Sect.~\ref{sec:pr_spe}, we observed a power-law component that is independent of the disk variability. Therefore, this component is unlikely driven by the inverse Compton scattering process between the disk photons and electrons in the corona. 
We therefore consider jets as the origin of the power-law component.
In the jet, particles could move perpendicularly away from the accretion disk at relativistic speeds within a small solid angle \citep[e.g.,][]{Pushkarev2009,Pushkarev2017,Tetarenko2017,Miller2019,Zdziarski2022}.This reduces the probability of particles in the jet scattering with disk photons, implying that photon propagation between the disk and jet is less efficient than between the disk and corona \citep[e.g.,][]{Wilkins2012}. 
This explains why the jet emission could vary independently from the disk emission in short timescales.

Furthermore, our multi-wavelength SED fitting also favors the jet origin, where the X-ray emission is dominated by SSC radiation. This requires a compact blob with a radius of $R = (0.1$--$2.5)\times10^8$\,cm and a strong magnetic field of $B = (0.3$--$3.5)\times10^6$\,G. This magnetic field strength is larger than those typically measured in most BHXRBs, which are generally on the order of $10^4$\,G, as seen in systems like GX~339--4 \citep{Shidatsu2011}, XTE~J1550--564 \citep{Chaty2011}, MAXI~J1836--194 \citep{Russell2014}, Cygnus~X--1 \citep{Zdziarski2014}, and MAXI~J1535--571 \citep{Russell2020}. However, studies of MAXI~J1820+070 \citep{Rodi2021,Echibur2024} and GRS~1915+105 \citep{Punsly2011} suggest that the launching region of the jet requires a magnetic field of $B\sim10^5$--$10^7$\,G, which is consistent with our results.

It is important to note that the decoupling between inverse Compton scattering from the corona and synchrotron+SSC radiation in the jet is model-dependent. The alternative scenario, in which the non-thermal X-rays are dominated by inverse Compton scattering from the corona, cannot be ruled out. Related analyses supporting this perspective have been conducted by \cite{Draghis2023} and \cite{Wang2024ApJ...963...14W}.

\subsection{Determining the distance to IGR~J17091} \label{sec:RX_plane}

Due to the unusual types of variability shared by GRS~1915 and IGR~J17091, it has been argued that the latter is a faint version of the former, either ascribed to a large distance or a smaller BH mass \citep[e.g.,][]{Altamirano2011,Wang2018MNRAS.478.4837W}. However, owing to the high extinction in front of IGR~J17091, its precise location and hence the central black hole mass are uncertain.

The logarithmic-linear relation between radio and X-ray emission has been verified in all types of accreting systems \citep[e.g.,][]{Merloni2003, Corbel2013}.
GRS~1915 is situated in the top right corner, beyond both the standard and hybrid radio–X-ray correlations for galactic BHs. Due to its extra-long outburst spanning over 26 years \citep{Neilsen2020}, this deviation has been attributed to a high mass accretion rate in GRS~1915.
While the similarities of GRS~1915 and IGR~J17091 suggest that the two targets may share the same slope in the radio–X-ray fundamental plane.

\begin{figure}[!t]
 \centering
   \includegraphics[width=1\linewidth]{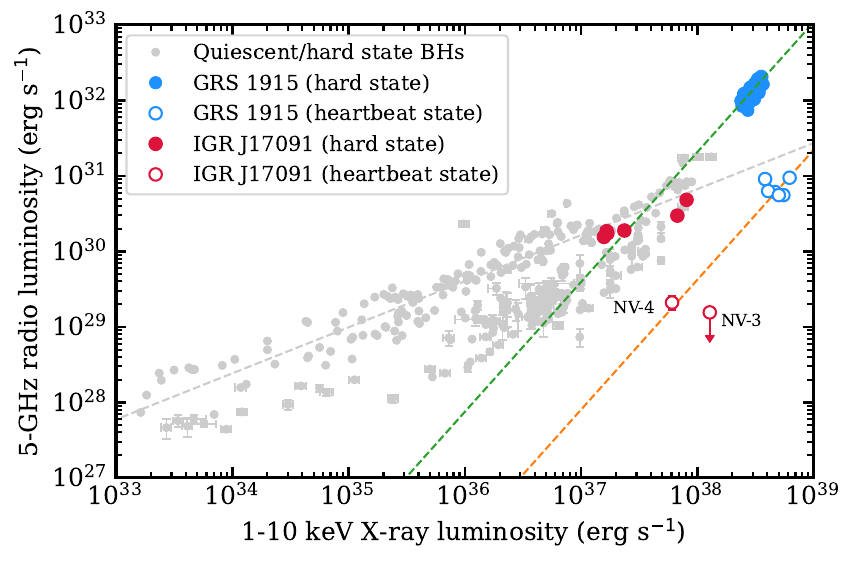}
   \caption{Radio–X-ray correlation for BHs. The grey points represent the data in quiescent/hard states.
   The grey dashed line has a slope of $\xi=0.61$ while the green and orange dashed lines have a slope of $\xi=1.72$. The distance of IGR~J17091 is assumed to be 13.7\,kpc.}
   \label{fig:RX_plane.png}
\end{figure}

We adopted the radio–X-ray fundamental plane for BHs from \cite{arash_bahramian_2022_7059313} and plotted as grey dots in Fig.~\ref{fig:RX_plane.png}. Additionally, we added both the radio and X-ray luminosities in the hard and heartbeat states of GRS~1915 as blue filled and open dots to the figure. The data in the hard state of GRS~1915 were adopted from \cite{Rushton2010}, while the radio data in heartbeat states were adopted from \cite{Klein2002}. Regarding the X-ray data in the heartbeat state, we fitted each available {\it RXTE} spectrum with the model $\texttt{tbabs}\times(\texttt{diskbb} + \texttt{powerlaw})$ to calculate the X-ray flux in the 1--10\,keV band.
For IGR~J17091, we adopted the radio and X-ray data in the hard state from \cite{Rodriguez2011, Gatuzz2020, Russell2022ATel}, while the data in the heartbeat state are from our study. To obtain the X-ray luminosities in the 1--10\,keV band, we fitted the \nicer spectra of IGR~J17091 with the same model outlined above. The radio data are from the ATCA 5.5\,GHz band.

As the data in the heartbeat state of GRS~1915 are rather scattered, we only measured the radio/X-ray slope, $\xi$, in the hard state and obtained $\xi=1.72\pm0.13$, and assume that its heartbeat state shares the same slope as the hard state. The correlation track in the hard (green) and the heartbeat (orange) states of GRS~1915 is shown in Fig.~\ref{fig:RX_plane.png}. We then applied these two correlation track to the data in the hard and the heartbeat states of IGR~J17091, respectively, and used the least-squares method to estimate its distance. We derived $D_{\rm hard} = 11.8^{+3.8}_{-2.9}$\,kpc for the hard state and $D_{\rm HB} = 15.1\pm2.8$\,kpc for the heartbeat state.
By incorporating these into a joint probability distribution, we obtained the best-fitting distance to IGR~J17091 to be $D=13.7\pm2.3$\,kpc. We adopt this distance to calculate the luminosity and plot the data of IGR~J17091 as red dots in Fig.~\ref{fig:RX_plane.png}.
However, the new type of variability, class X, in NV-3 deviates from this relationship. This suggests that the emergence of this new type may require additional physical conditions to be met for activation.

\section{CONCLUSIONS} \label{sec:CONCLUSIONS}

In this work, we use archival data from \nustar, \nicer, \swift, \fermi, \neowise, and \atca to study the origin of the non-thermal X-ray emission in the heartbeat state during the 2022 outburst of IGR~J17091. We find that the short-term variability arising from the accretion disk plays a crucial role in assessing and decoupling the contributions of the corona and the jet in the X-ray band. 

Regarding the newly identified type of variability, Class X, although it presents a lightcurve profile opposite to that of Class IV, both can be driven by radiation pressure instability in the accretion disk. Moreover, we observed a significant component in the lag-frequency spectrum of Class X. This component exhibits different evolutions in both the rms and lag energy spectra compared to Class X, suggesting it may involve a separate emission region contributing to another form of quasi-periodic variability in X-rays.

Moreover, we suggest the distance to IGR~J17091 of $D=13.7\pm2.3$\,kpc by assuming it shares the same radio-X-ray fundamental plane relationship as GRS~1915 during both the hard state and heartbeat states.

In X-ray timing and spectral analysis, we observe a power-law component that is independent of the heartbeat-like variability. Further quasi-simultaneous broadband SED analysis suggests that this power-law component in the X-ray can be explained by SSC radiation within a compact blob, sized of $R = (0.1$--$2.5)\times10^8$\,cm. This requires a strong magnetic field of $B = (0.3$--$3.5)\times10^6$\,G. In this case, our SED fitting suggests that the radio emission originates from a different region than the X-ray jet, consistent with the multi-zone jet model \citep[e.g.,][]{Kaiser2006, Kantzas2021, Lucchini2022, Tramacere2022}. However, the sparsity of our data makes a more detailed study of jet behavior difficult to establish. 

Future simultaneous observations from infrared to radio bands are essential to refine our findings and further constrain the jet parameters \citep[e.g.,][]{Rodi2021, Echibur2024}. Observations at higher radio frequencies above 100\,GHz (ALMA) can further constrain the spectral index of the radio spectrum and determine the cooling break of the synchrotron emission \citep[e.g.,][]{Russell2014,Tetarenko2015}, whereas high-sensitivity observations in the infrared (VLT and \textit{JWST}) can be used to solve the model degeneracy in the synchrotron component of the compact jet region \citep[e.g.,][]{Rodi2021}. Additionally, high-sensitivity radio observations below 1\,GHz (MeerKAT and SKA) would help determine the transition frequency in the radio band between the optically thin synchrotron emission region and the synchrotron self-absorption region \citep[e.g.,][]{Nakar2011}. These measurements can help constrain the multi-zone jet parameters and further establish the overall jet structure.

\begin{acknowledgments}

We thank Liang Chen, Defu Bu, Erlin Qiao, Shu Wang, and Zeyang Pan for the useful discussion, and Jakob van den Eijnden for the help with the optimal filtering algorithm. We thank the anonymous referee for the comments.
This research was supported by the National Natural Science Foundation of China (NSFC) under grant numbers 11988101, 11933004, 12173103, and 12261141691; by the National Key Research and Development Program of China (NKRDPC) under grant numbers 2019YFA0405504 and 2019YFA0405000; and by the Strategic Priority Program of the Chinese Academy of Sciences under grant numbers XDB41000000 and XDB0550203. SdP gratefully acknowledges support from the ERC AdvancedGrant 789410.

\end{acknowledgments}

\bibliography{main}{}
\bibliographystyle{aasjournal}

\appendix
\renewcommand{\thefigure}{\hbAppendixPrefix\arabic{figure}}
\numberwithin{figure}{section}
\section{MCMC CORNER PLOTS of SED fitting} \label{sec:MCMC}
In this section, we present the corner plots of the SED fitting parameters from the MCMC for cases (a), (b), and (c) described in Sect.~\ref{sec:SED}. These plots show the significant degeneracy between the fitting parameters.

\begin{figure*}[!h]
\centering
\mbox{\includegraphics[width=0.32\linewidth]{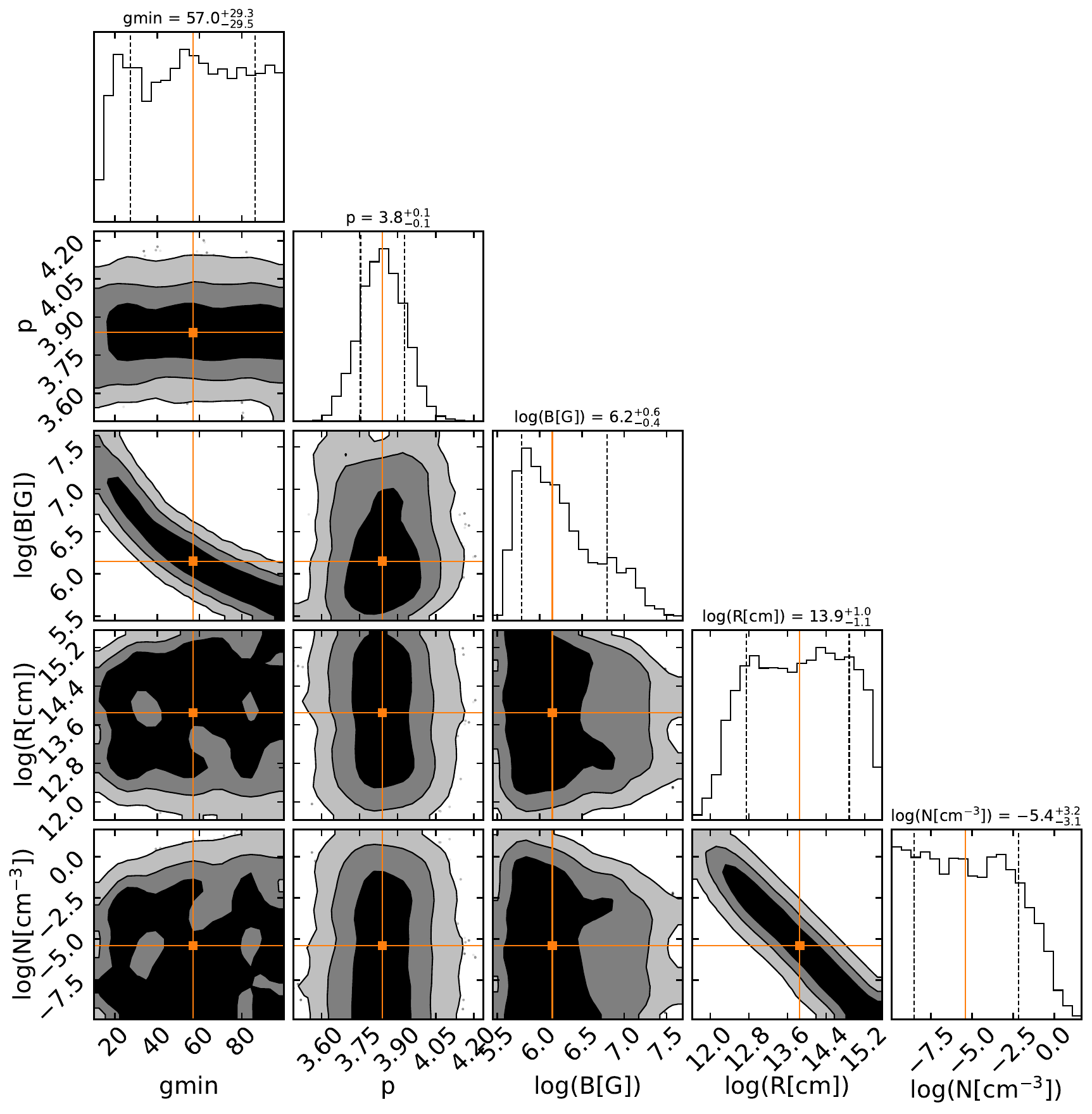}}
\mbox{\includegraphics[width=0.32\linewidth]{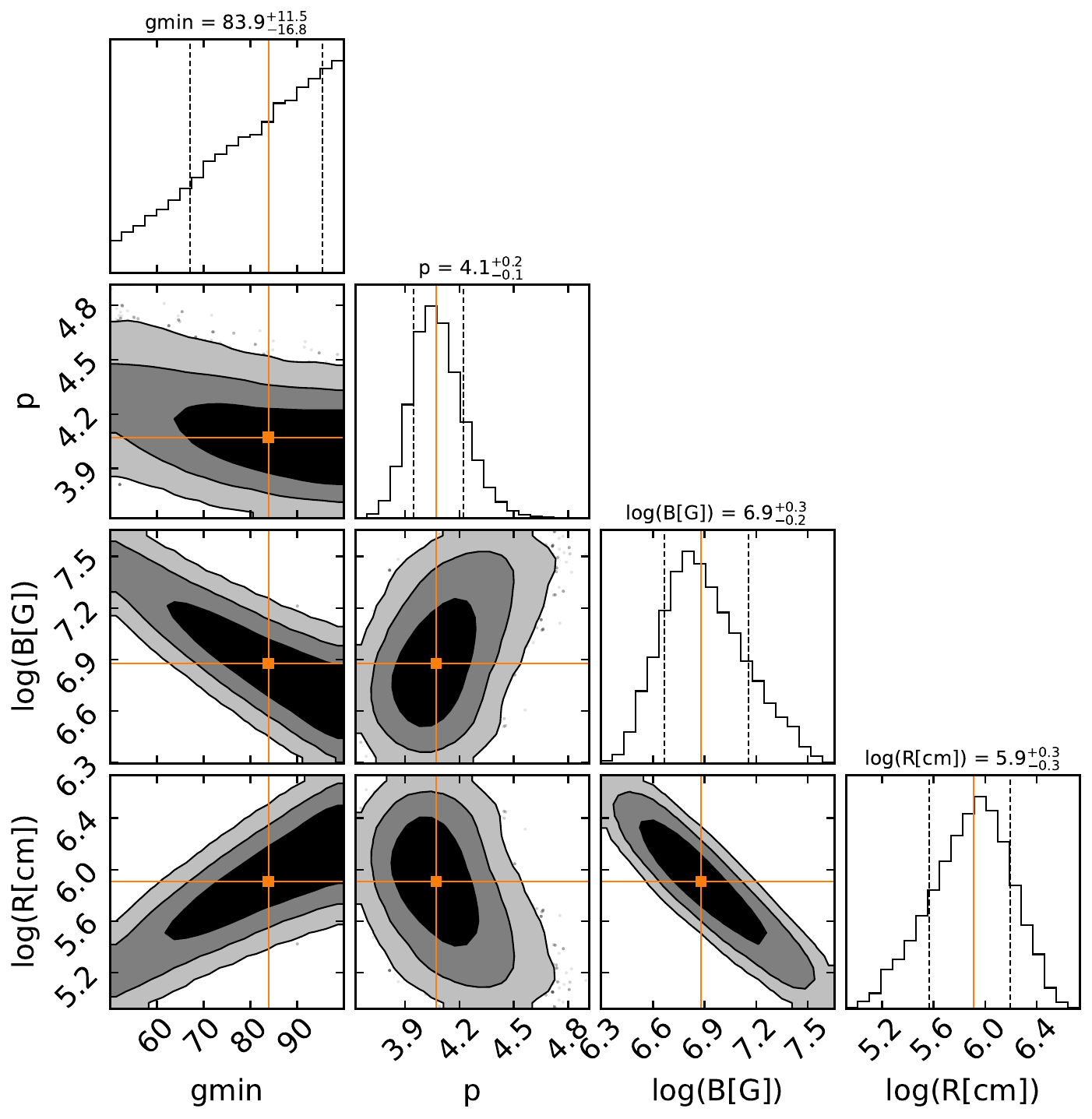}}
\mbox{\includegraphics[width=0.32\linewidth]{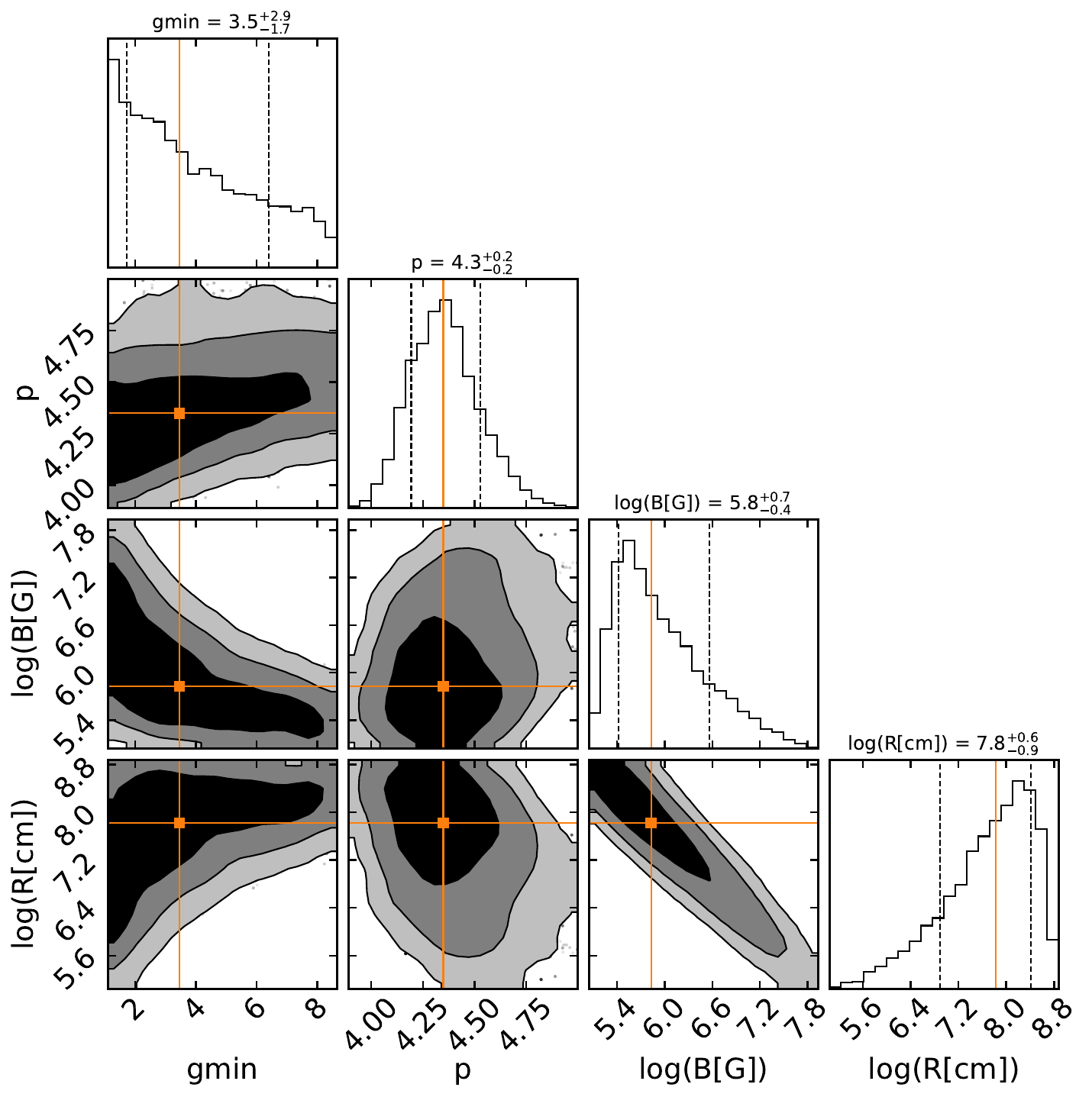}}
\caption{The corner plots of the SED fitting parameters, from left to right, correspond to cases (a), (b), and (c) described in Sect.~\ref{sec:SED}. The contours contain the 1, 2, and 3-$\sigma$ confidence intervals respectively. The central orange line represents the median value of each parameter from the MCMC chain, with two black dashed lines on either side showing the 1-$\sigma$ confidence interval.
}
\label{fig:MCMC_c.png}
\end{figure*}

\end{document}